# Premio Nobel de Física 2020
# Hoyos negros y sus misteriosos interiores

## José M. M. Senovilla

Los teoremas de singularidades constituyen uno de los mayores hitos de la relatividad. Generaron una panoplia de fértiles líneas de investigación con consecuencias físicas deslumbrantes.

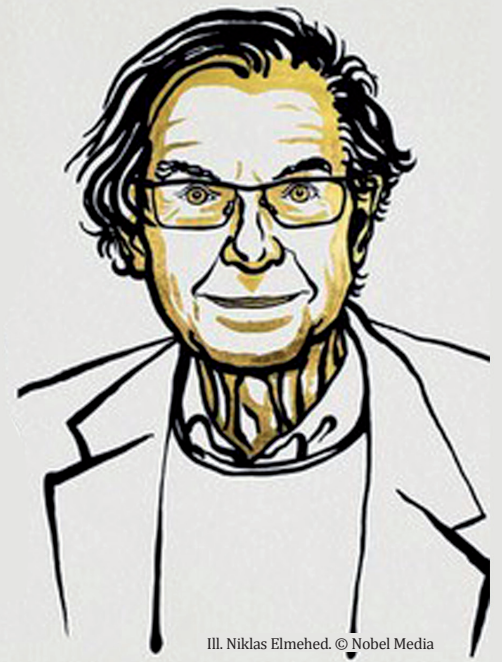

Ill. Niklas Elmehed. © Nobel Media

El premio Nobel de 2020 ha recaído en la física de los hoyos, o agujeros, negros. La mitad del galardón ha laureado a sir Roger Penrose por ser el fundador de la teoría subyacente y por haber demostrado, en particular, un teorema de singularidades [1] que cambió el rumbo de la relatividad general (RG) para siempre. *El teorema de Penrose nos informa del interior de los hoyos negros*. La RG alcanzó su madurez con estos resultados. Emergió una teoría renovada, vigorosa, plagada de sorpresas. Penrose merece el premio por una carrera original y genial, por haber dotado a la RG de herramientas valiosísimas, por conceptos clave tales como el de esferas atrapadas y por su influencia determinante y arrolladora.

### Antecedentes

Para comprender el alcance y la importancia de los resultados laureados es conveniente hacer una semblanza histórica de los principales avances previos.

El 25 de noviembre de 1915 Einstein presenta las ecuaciones de su teoría de la RG. En ella se unifica el espacio con el tiempo entreverándose además con la geometría y la materia-energía. Cualquier cosa que exista tiene energía y por ello gravita: su campo de gravedad se manifiesta en la geometría del espacio-tiempo circundante. Apenas un mes después Karl Schwarzschild encontró la solución exacta para el campo gravitatorio exterior de cualquier distribución material con simetría esférica. Aparecía un infinito en r = 0, una *singularidad*, donde r es una variable tal que las esferas seleccionadas por la simetría tienen área $4\pi r^2$. No pareció preocupante, se podía imaginar que es similar a la divergencia que aparece en r = 0 para el campo newtoniano central cuya intensidad es proporcional a $1/r^2$. *A posteriori* esta idea intuitiva se demostró engañosa (ver explicación en Figura 2).

La solución de Schwarzschild tenía *otro problema* en r = $r_g$= $2GM/c^2$, denominado *radio gravitatorio* (a menudo radio de Schwarzschild), con G la constante de la gravitación universal, c la velocidad de la luz en el vacío y M la masa total del cuerpo esférico que crea el campo gravitatorio. Preocupado por este inesperado inconveniente Schwarzschild resolvió las ecuaciones en el interior de la materia suponiendo por simplicidad un fluido incompresible (densidad de masa constante). Comprobó que las dos soluciones, interna y externa, podían juntarse armoniosamente solamente si el valor de r en la superficie de la esfera material era apreciablemente mayor que $r_g$. Concluyó que valores de r < $r_g$ eran inaccesibles en situaciones reales. Para entender la magnitud de $r_g$ repararemos en que su valor es de 9 mm para la masa de la Tierra, o de $10^{-25}$ m para una persona de 75 kg (¡tan puntual quizás como un quark!).

En 1922-24 Fridman encontró las soluciones *dinámicas* de las ecuaciones que representan los modelos en expansión que hoy describen la evolución del Universo a gran escala. Contenían un instante de tiempo pasado donde la densidad de masa se hacía infinita y el espacio se quebraba, literalmente desaparecía: lo llamó "el instante de la creación". Esto inquietó a Lemaître, quien en 1933 se planteó resolver la siguiente discordancia: en los modelos de Fridman las esferas pueden tener áreas tan diminutas como se desee, lo cual es incoherente si existe un límite impuesto por el radio gravitatorio $r_g$. Demostró que el límite de $r_g$ es ilusorio, consecuencia de haber supuesto una solución independiente del tiempo. Mostró cómo eliminarlo con un simple *cambio de coordenadas* que *extiende* la solución para valores de r < $r_g$, valor éste que describe un *horizonte regular*. Este tipo de horizontes ya están presentes en el espacio-tiempo de Minkowski, vale decir en relatividad especial, para ciertos observadores. Hay una explicación en la Figura 1.

Surgió entonces una pregunta de relevancia física: ¿son esas regiones con r < $r_g$ físicamente accesibles para estrellas u otros objetos cósmicos realistas?

### Oppenheimer y Snyder: los hoyos negros

A caballo entre las décadas de 1920-30, Anderson, Stoner y Chandrasekhar descubrieron de manera independiente que las estrellas de tipo enana blanca no se estabilizan si los valores finales de la masa en su proceso de nacimiento son mayores que un cierto límite, que hoy se estima en 1.4 $M_\odot$. Posteriormente, inspirados en las estrellas de neutrones predichas por Landau, Baade y Zwicky a principios de los treinta (aunque no fueron detectadas hasta 1967 por Jocelyn Bell), Oppenheimer y su discípulo Volkoff examinaron en 1939 su estabilidad teniendo en cuenta los efectos de la RG. Encontraron otro límite superior de masa a partir del cual la estabilidad final era imposible. Teniendo en cuenta las fuerzas nucleares, hoy en día se tasa el límite en el rango 2.2-2.9 $M_\odot$. *La gravedad siempre acaba imponiéndose si hay masa suficiente.*

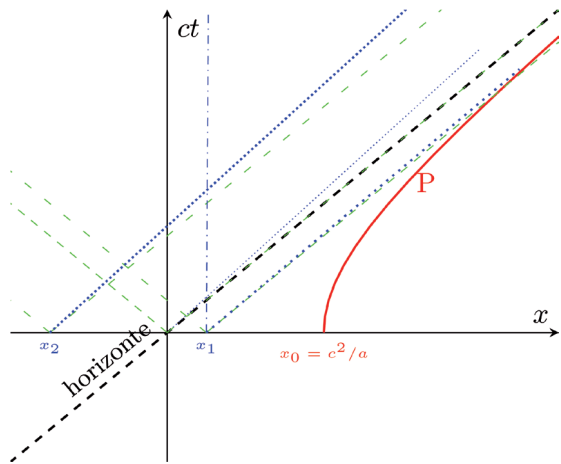

Oppenheimer, junto con Snyder, consideró entonces el colapso de materia esférica pulverulenta (es decir, sin presión) en RG. Su artículo [2] es un clásico que aún en el presente proporciona la imagen básica de cómo discurre el proceso de colapso imparable y continuado sin alcanzar el equilibrio. Mostraron cómo la estrella tiende a "sellarse a sí misma" y una vez cruza su radio gravitatorio —lo cual sucede en un tiempo *propio* finito— se desconecta causalmente, de manera completa, de cualquier observador externo. Permanece, eso sí, el campo gravitatorio del astro que se derrumba.

La percepción que tiene una observadora externa de cómo transcurre el colapso es radicalmente diferente (Figura 2). Jamás podrá observar que la estrella atraviesa su radio gravitatorio que es, de hecho, un *horizonte* para ella: el tiempo que transcurre hasta el establecimiento de la ocultación del astro es para ella infinito. Aunque la situación es excesivamente idealizada, se argumenta en [2] que las conclusiones no cambiarán en exceso considerando desviaciones de la esfericidad y/o valores no nulos de la presión. Quedaba así inaugurada la aventura de los hoyos negros (HN).

### Los teoremas previos

Apenas un mes después del fallecimiento de Einstein (18 de abril de 1955) Raychaudhuri publicó un resultado clave [3] y, poco después, Komar [4] de manera independiente otro similar, pero más general. Estas referencias contienen el primer *teorema de singularidades* de la historia y la *ecuación de Raychaudhuri*, base de todos los teoremas posteriores.

Dicha ecuación es una relación geométrica que relaciona la evolución de la divergencia de un campo vectorial a lo largo de su flujo con (i) sus propiedades cinemáticas, tales como la cizalladura, la rotación y la aceleración y (ii) el campo gravitatorio. Se ve todos los términos incrementan el poder atractivo del campo de gravedad excepto la rotación y quizás la aceleración, produciendo un efecto netamente focalizante, en el sentido de que las curvas integrales del campo vectorial tienden a confluir. La atracción gravitatoria se manifiesta en la positividad de una cierta combinación lineal de la densidad de energía y las presiones reinantes. Así es como Komar introdujo por primera vez lo que luego se ha dado en llamar *condición de energía*, o *condición de convergencia*, que con el tiempo se convertiría en la primera suposición corriente de todos los teoremas de singularidades posteriores.

En [3] y [4] se supuso que el campo vectorial describía el movimiento de un fluido real sin rotación ni aceleración, de forma que según la ecuación nada se oponía al efecto focalizador. Se deducía entonces que si el fluido comenzaba a contraerse (o expandirse) alcanzaría (provendría de) una singularidad fatídica donde la densidad de masa diverge en un tiempo finito. No obstante, la ecuación de Raychaudhuri es válida para cualquier conjunto continuo de curvas independientemente de que éstas describan el movimiento de un fluido o no. En el caso de curvas abstractas se produce una "cáustica" de la familia de curvas, o sea, un lugar donde las curvas se cruzan, análogamente al foco de una lente donde convergen los rayos luminosos. A este fenómeno universal, válido si la condición de convergencia se satisface, se le denomina *efecto de focalización* de la gravedad.

### El descubrimiento de los cuásares

El final de la guerra produjo un cambio radical en la astronomía, que pasó a usar los avances en la manipulación y detección de ondas de radio para observar los cielos en tales frecuencias. Un hecho de especial relevancia fue la identificación de la radio-fuente 3C 273 como extra-galáctica por Maarten Schmidt en 1962. El resultado era "increíble", porque esa fuente debía estar a una distancia de aproximadamente 2 500 millones de años-luz, parecía ser puntual, por lo que debía ser muy compacta, y emitía un cantidad de energía gigantesca, comparable o mayor que la de ¡galaxias enteras!

En seguida se fueron acumulando las observaciones de este tipo de fuentes, que se dieron en llamar "objetos cuasi-estelares", luego abreviado a *cuásares*. Son un tipo de *núcleo activo de galaxia*, cuyo mecanismo de producción de energía actualmente se cree es el enorme barullo que provoca un HN superlativo a su alrededor.

En aquella época, sin embargo, no se conocía ningún procedimiento capaz de originar tales potencias. Las reacciones nucleares eran una respuesta harto improbable, porque su eficiencia para producir ondas electromagnéticas a partir de materia es insuficiente. Sin embargo, la radiación emitida por partículas aceleradas en objetos hipercompactos de muchísima masa parecía ser un candidato adecuado. En fin, los cuásares causaron revuelo considerable y la agitación aumentó en 1963 con el descubrimiento por Roy Kerr de la solución de las ecuaciones de Einstein que hoy sabemos describe el campo de un HN en rotación. En seguida se comprendió que un disco de acrecimiento en torno de un HN de Kerr podía aumentar la eficiencia para convertir materia en energía desde un nada despreciable 6 % (para el caso sin rotación de Schwarzschild) hasta aproximadamente un 42 % si la rotación del hoyo es cercana a la máxima permitida. ¡Qué barbaridad!

Ese mismo año se celebró en Dallas el Texas Simposium on Relativistic Astrophysics —organizado por cierto por relativistas matemáticos— dedicado al entonces reciente descubrimiento de los cuásares y su relación con la RG. Kerr presentó allí su nueva solución, aunque poca gente pareció hacerle mucho caso... [5].

### El teorema de Penrose (1965)

John A. Wheeler se propuso clarificar el problema de las singularidades. Discutió con Penrose la importancia de resolver esta cuestión, en particular para confirmar, o no, si todo se debía a las idealizaciones exageradas y poco realistas de las soluciones conocidas. Tanto el **modelo de Oppenheimer-Snyder** como el teorema de Raychadhuri-Komar ignoran los efectos de la presión y de la rotación y, además, aquél tiene simetría esférica exacta. Penrose, por su parte, conocía perfectamente la solución de Kerr y su estructura causal, y en particular la existencia de una singularidad; pero al fin y al cabo la métrica de Kerr retenía demasiada simetría.

En 1963 E. M. **Lifshitz e I. M. Khalatnikov** atacaron por su parte el problema, llegando a la conclusión de que las singularidades no eran genéricas, sino que más bien lo usual sería la aparición de cáusticas, problema menor resoluble por adecuada elección de coordenadas. Penrose dudó del resultado y las matemáticas subyacentes.

En diciembre de 1964, A. G. Doroshkevich, Ya. B. Zel'dovich e I. D. Novikov enviaron un artículo (publicado en 1965) donde se argumentaba que los procesos **físicos acababan eliminando las irregularidades no elásticas en un colapso estelar y devenían ignorables, lo que reforzaba una descripción similar a la de Oppenheimer-Snyder (Figura 2). Reafirmaron en particular la desconexión causal de la estrella respecto de los observadores externos. No dejaron de preguntarse, no obstante, qué podía ocurrir dentro, más allá del radio gravitatorio una vez superado éste por la estrella —o lo que quede de ella.

Pertrechado con la ecuación de Raychaudhuri, la condición de convergencia y su sabiduría e imaginación, Penrose apareció en escena con un brevísimo artículo [1] (dos páginas y poco) que cambió el curso de la RG para siempre. Los matemáticos,

como él, sólo prueban teoremas usando conceptos bien definidos. Penrose combinó tres hipótesis:
- la condición de convergencia,
- existencia de una hipersuperficie de Cauchy no compacta,
- existencia de una *esfera atrapada*.

La conclusión era diáfana: es imposible que todos los rayos de luz sean *completos*, o sea, que puedan viajar hacia el futuro indefinidamente.

Aquí hay dos ideas profundamente transformadoras, una de ellas genial (esfera atrapada), la otra innovadora (incompletitud). Antes de entrar en el detalle de estos dos conceptos fundamentales, aclaremos que una hipersuperficie de Cauchy es un instante de tiempo apropiado para dar datos

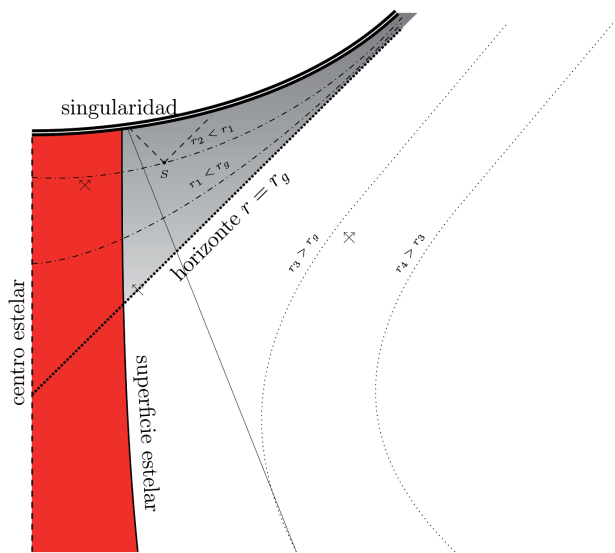

**Figura 2.** Esquema de la formación de un hoyo negro por colapso de una estrella esférica representada en la zona roja. El tiempo transcurre en la dirección vertical y la luz radial, como en la Figura previa, viaja en rayos a 45° como la vertical representados por las cruces flechadas. La idea más fiel se obtiene si uno hace rotar el dibujo en torno al eje definido por el centro estelar. Cada punto (fuera del centro) dibuja entonces una circunferencia que representa una esfera real. Estas esferas tienen área $4\pi r^2$, y las líneas de puntos o de puntos y rayas indican zonas de área constante para cada valor de $r$. Un cuerpo en órbita de la estrella/HN, con periastro y apoastro en $r_1$ y $r_4$ respectivamente, tiene una línea de universo contenida entre las dos líneas de puntos representadas. Ese cuerpo, o cualquier otro que permanezca en la zona con $r > r_g$, sólo es capaz de observar/sentir la estrella antes de que cruce el horizonte $r = r_g$. Nótese que desde el exterior el campo gravitatorio que se experimenta es *siempre el campo de la estrella primogenitora antes de que se convierta en un HN*. Cuando la superficie estelar cruza el valor $r_g$ se desconecta de la parte exterior del universo y se encierra detrás de un horizonte análogo al de la Figura previa para P. El interior del HN, representado por la zona sombreada, tampoco puede comunicarse con el exterior. Eso sí, uno puede dejarse caer radialmente hacia el HN (línea continua) y, al atravesar el horizonte, pasar a formar parte de la zona íntima del HN invisible desde el exterior. Las esferas internas con valores de $r < r_g$, tales como S, son *atrapadas* (Figura 3): es fácil observar que, independientemente de cómo se viaje hacia el futuro a partir de S, su área decrece irremediablemente, hasta llegar a la singularidad, donde el área se anula. Obsérvese que la singularidad está *en el futuro* de cualquier observador y por ello es totalmente invisible, *tanto para los observadores externos como para los que entran en el HN*.

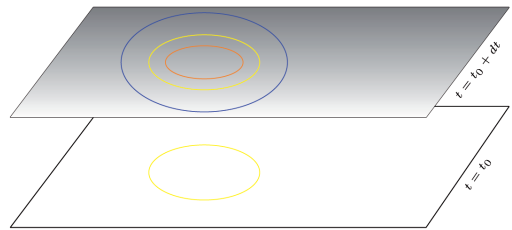

Situación estacionaria

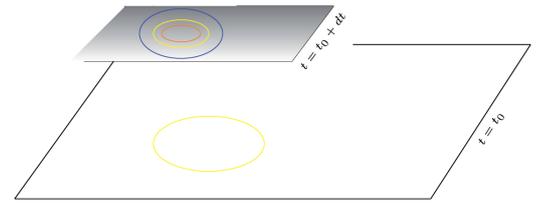

Espaciotiempo en contracción

iniciales de las ecuaciones de Einstein. Para entonces ya se sabía, debido al trabajo de Fourès-Bruhat (luego Choquet-Bruhat), que las ecuaciones de la RG son hiperbólicas, lo que quiere decir que matemáticamente los problemas de condiciones iniciales están bien definidos y, para datos iniciales adecuados, la solución evoluciona convenientemente, existe y es única. Una hipersuperficie es "buena" para dar datos iniciales si cualquier línea de universo posible (temporal o luminosa) la cruza antes o después. A estas hipersuperficies se las denomina "de Cauchy". Por ejemplo, un hiperplano t = constante en el espacio-tiempo de Minkowski lo es. Penrose supuso la existencia de tal tipo de hipersuperficie con el añadido "no compacta" (como el hiperplano mencionado lo es). De alguna manera se está diciendo que el espacio es infinito.

### Incompletitud geodésica

En RG las partículas en caída libre siguen curvas *geodésicas* que por definición hacen máximo el intervalo espacio-temporal entre dos puntos. Si la partícula tiene masa en reposo, la geodésica es temporal y maximiza el tiempo; si no, como los fotones, la geodésica es de género luz, o luminosa, y el intervalo es simplemente cero.

Una geodésica se dice *completa* si se puede extender para valores arbitrarios de su parámetro canónico (que es el tiempo propio para las temporales). La idea innovadora de Penrose consistió en probar la existencia de geodésicas luminosas incompletas, que representan rayos de luz abortados, desaparecidos. Estos rayos malogrados indican una incompletitud del espacio-tiempo que es señal inequívoca de la existencia de la *singularidad.* Porque dar una definición rigurosa de singularidad en RG es un cometido peliagudo [6]. En el caso arriba mencionado del campo central newtoniano, su intensidad diverge, pero el espacio-tiempo queda indemne y se puede decir la posición-instante de la singularidad. En el caso de singularidades gravitatorias, empero, el espacio-tiempo mismo falla, y por esto las singularidades *no están en él.* Aquí brilla la originalidad de Penrose: es oportuno diagnosticar la presencia de una singularidad si hay caminos que se acercan, pero cuya semblanza se acaba. En ausencia de **singularidades** estos caminos seguirían indefinidamente.

En resumidas cuentas, si algo proviene de una singularidad pasada se nos muestra como si surgiera *ex nihilo* (instante de la creación); si algo se acerca a una singularidad futura se nos manifiesta como si súbitamente se esfumara.

### Esferas atrapadas

Como bien se sabe, es muy difícil escapar de un campo de gravedad. Ya en la Tierra hace falta un montón de energía para, digamos, llegar a la Luna (que todavía está en su campo de influencia). En física clásica se usa el concepto de velocidad de escape, pero en RG las cosas son bastante más complicadas.

Hasta que llegó Penrose, claro está, y nos enseñó el concepto de *esfera atrapada,* su idea genial. Recordemos que la gravedad es la geometría del espacio-tiempo. En situaciones dinámicas la geometría evoluciona con el tiempo, así que áreas, volúmenes, etcétera son magnitudes cambiantes con el tiempo según lo que dicte el campo gravitatorio. Perfectamente puede suceder que el área de algunas esferas disminuya con el tiempo si el mundo está en contracción. Tales esferas ofrecen a todo lo que envuelven menos metros cuadrados para evadirse. Aún así, uno puede decidir escapar simplemente corriendo hacia afuera de la esfera tan rápidamente como se pueda, incluso los fotones con la velocidad de la luz. En situaciones de gravedad extrema resulta que, sorprendentemente, es posible que, aun intentando escapar, a medida que transcurre el tiempo uno se encuentre constreñido por esferas de área *aún menor* (Figura 3).

La noción de esfera atrapada es una contribución trascendental a la RG, a la física gravitatoria en general y a la geometría. Para empezar, es un concepto independiente de la existencia de simetrías. De mayor calado, las esferas atrapadas son *estables* frente a perturbaciones. Ya se sabía que este tipo de esferas existen en la solución de Kerr, y también en la zona interna del modelo de Oppenheimer-Snyder (Figura 2). La estabilidad indica por lo tanto que también existirán en otros colapsos sin simetría o con materia más realista pero idealizada por esos modelos. En plata: ni las desviaciones de la simetría esférica ni otras consideraciones físicas son capaces de prevenir la formación de singularidades *una vez aparecen las esferas atrapadas*.

Otro teorema [7] asegura que ninguna esfera atrapada se asoma fuera del horizonte de sucesos de un HN. En consecuencia, el teorema de Penrose ¡nos informa de lo que pasa en el interior de un HN! Allá dentro (a) hay esferas atrapadas y, por ello, (b) el espacio-tiempo es incompleto, *singular*.

O al menos eso es lo que predice la relatividad general.

### Los teoremas de singularidades

El teorema de Penrose impresionó a la comunidad relativista, con entusiasmo para unos y perturbación para otros. Su impacto fue instantáneo y muy hondo. En seguida se desarrollaron una plétora de teoremas de singularidades, principalmente por Hawking y Geroch. En particular, Hawking fue el primero en percatarse de que los modelos de Fridman contienen **esferas atrapadas (al pasado)**, y demostró varios teoremas de aplicación cosmol**ógica. En 1970 Penrose y Hawking [8]** aunaron esfuerzos e ideas para probar el considerado teorema de singularidades por antonomasia.

Con el tiempo se han ido probando numerosos teoremas de singularidades, afinándolos en múltiples direcciones. Todos ellos tienen la misma estructura básica, como sugerí en su momento [9]. El "teorema patrón" contiene tres ingredientes y una conclusión del siguiente tenor [9, 10]:

Si el espacio-tiempo satisface
1. condiciones de convergencia,
2. un buen principio de causalidad,
3. y una adecuada condición de contorno/inicial
entonces es incompleto.

Con esta estructura se han logrado encontrar alternativas cada vez menos exigentes para las tres suposiciones 1-3. Los teoremas se han ido acomodando a diversas novedades teóricas —dimensiones extra, inflación cósmica, radiación de Hawking, etc— y a descubrimientos observacionales —aceleración de la expansión universal—. Una línea de especial relevancia trata de los teoremas que buscan incorporar efectos cuánticos dada nuestra ignorancia actual acerca de la gravedad cuántica. En fin, sería farragoso y pesado enumerar aquí todas las vertientes de los teoremas de singularidades, el lector interesado puede consultar [10].

### El legado de Penrose

Sir Roger Penrose es un gran matemático que ha influido en varias ramas de álgebra (inversa matricial de Moore-Penrose), geometría (teselados no periódicos del plano), ilusiones ópticas (donde colaboró con Escher, el artista), sistemas integrables y teoría de representaciones (teoría de los "twistors"), además de en relatividad, tanto especial (el efecto Penrose-Terrell) como general. La lista está lejos de ser completa. El premio Nobel se otorga (en física, no así en paz o literatura) por trabajos concretos, mas si se concediera por carreras completas no es fácil mencionar un aspirante de su talla.

Aquí me quiero concentrar, obviamente, en su legado relativista. Aparte de lo ya dicho, es necesario resaltar que su breve artículo de 1965, por el que seguramente será principalmente recordado a partir de ahora, es una síntesis que contiene explícita o implícitamente, en un puñado de palabras, muchas de sus ideas y conceptos desarrollados en la década de los sesenta y principios de los setenta, y que a la postre resultaron en la era moderna de la RG, su versión puramente posteinsteinana [11, 10].

Su bautizo en RG comienza el año de mi nacimiento, con su artículo original sobre el uso de espinores (de dos componentes complejas) [12], una idea brillante para la relatividad 4-dimensional basada en las representaciones espinoriales del grupo de Lorentz. Este artículo subyace su obra posterior y queda plasmada en dos volúmenes densos y fundamentales (junto con Rindler) [13]. Entre otros muchos resultados originados en esta idea aparecen las ecuaciones, o formalismo, de Newman-Penrose [14] que en particular contiene la "ecuación de Raychaudhuri" para curvas luminosas, luego usada en su teorema del Nobel.

El siguiente paso de gigante fue su brillante idea de estudiar las propiedades asintóticas de los espacio-tiempos, en especial de los campos radiativos, usando técnicas de compactificación conforme. Fue cuando aprendimos que la parte asintótica de un espacio-tiempo, su infinito, consiste en cinco partes (diferentes en general) conocidas como $i^+$, $i^-$, $i^0$, $\mathscr{I}^+$ y $\mathscr{I}^-$ [la última letra se pronuncia "scri", un juego de palabras entre *sky* (cielo) y *script-i* ("i" tipográfica)], así como la definición de espacio-tiempo *asintóticamente plano, o asintóticamente (anti)-de Sitter* [15, 16, 17]. Aquí residen los fundamentos de los hoy indispensables *diagramas de Penrose*, ubicuos en las publicaciones de RG y cosmología, herramienta básica para la comunicación entre relativistas (ver Figura 4 para un ejemplo).

Penrose fue el principal impulsor de la *teoría de la causalidad*. Comenzó probando que algo tan básico como el campo gravitatorio de una onda plana electromagnética es un espacio-tiempo (geodésicamente) completo, pero ¡no posee ninguna hipersuperficie de Cauchy! [18]. Este hecho estaba claramente relacionado con el poder de focalización de la gravedad actuando sobre sí misma (por la no linealidad de las ecuaciones). Propuso luego un estudio abstracto de espacios causales [19] para finalmente dar la definición de *borde* *causal* del espacio-tiempo [20], un concepto que ha tenido largo recorrido tanto en matemáticas como en física del siglo XXI (ya que tiene relevancia, por ejemplo, para la conjetura AdS-CFT).

Otro tema de gran importancia que inició fue el estudio de la dispersión y colisión de ondas gravitatorias con un artículo clave [21], en cuya sinopsis se planteaba la importancia de su detección para la astronomía observacional del futuro, ya que se podía obtener información significativa de procesos extremos que tengan lugar en el Universo. ¡Francamente visionario! En [21] se obtiene la primera solución para colisión de ondas planas viajando en el espacio-tiempo plano y se comprueba la producción de una singularidad debido al mutuo poder focalizador de las ondas. Esta singularidad se produce en el puro vacío, sin ningún tipo de materia por ningún lado, sólo las dos ondas interfiriendo. Aunque la suposición de ondas planas es irreal, trabajos más recientes de Christodoulou han probado la formación de esferas atrapadas en el puro vacío debido a la llegada de ondas gravitatorias de intensidad y extensión finitas.

Y para acabar este incompleto repaso del legado inabarcable de Penrose naturalmente hay que mencionar su fomento al desarrollo de todo lo relacionado con los hoyos negros [22, 23, 24], en particular (1) el *proceso de extracción de energía* de un HN [22, 25], (2) el concepto de *horizonte de eventos* [22, 23], y (3) su *hipótesis de la censura cósmica* [22, 26].

(1) El proceso para robar energía de un HN (necesariamente en rotación) es un procedimiento relativamente complicado para reducir el momento angular del HN usando cuerpos que se dividen en dos muy cerca del horizonte —allá donde la rotación del espacio-tiempo hace que las partículas sean arrastradas irremediablemente— de forma que una mitad se adentre en el HN, haciéndolo perder parte de su momento angular, mientras que la otra mitad escapa hacia afuera con mayor energía total que la inicial del cuerpo. Este proceso no viola el teorema de que el área (entropía) del horizonte de un HN nunca puede decrecer [24], ya que la fórmula para el área —en el caso con rotación— involucra tanto la masa del HN como su momento angular, y muy bien puede decrecer éste mientras el área crece. Una versión sofisticada de esta extracción de energía, denominado proceso de Blandford–Znajek, que usa campos electromagnéticos en las cercanías de un HN en rotación, es la explicación más aceptada hoy en día de los gigantescos chorros astrofísicos que se observan en HN supermasivos de centros galácticos, así como uno de los "motores" de los cuásares.

(2) Se ha explicado antes que los hoyos negros son los objetos más pudorosos que existen, se encierran en sí mismos y no permiten a ningún observador externo observar lo que contienen. La frontera que los separa del mundo exterior es su horizonte de eventos, que Penrose definió [22, 23], usando su concepto de infinito, como el borde de la zona de las partículas que serían capaces de escapar llegando a $\mathscr{I}^+$ (Figura 4). Es una frontera inmaterial, una zona (totalmente regular) del espacio-tiempo. Si alguna partícula intrépida se aventura a atravesarla para indagar en el interior del HN, pasará a formar parte de ello y jamás podrá salir (salvo quizás diluido en una radiación de Hawking $2 \times 10^{67}$ años después para un modesto HN de una masa solar). Esto implica que la singularidad interior predicha por los teoremas queda oculta, invisible, sin efecto alguno sobre el mundo externo. Lo contrario sería una catástrofe porque el colapso imparable conduciría a una *singularidad desnuda* [25] (Figura 5), lo que provocaría problemas irresolubles de predictibilidad en la teoría —aunque, todo hay que decirlo, la existencia de tales singularidades podrían ser la mejor ayuda para los exploradores de la gravedad cuántica—. En cualquier caso, como Penrose puso claramente de manifiesto [22], hay una diferencia esencial entre el *status* lógico del colapso que lleva a un HN con su singularidad interior y el caso de un colapso hacia una singularidad desnuda: en el primer caso hay esferas atra**padas, y el teorema de 1965 im**plica que aún con perturbaciones genéricas de los modelos conocidos habrá una singularidad. En el segundo no hay esferas atrapadas y, por tanto, nada que asegure la estabilidad de la singularidad.

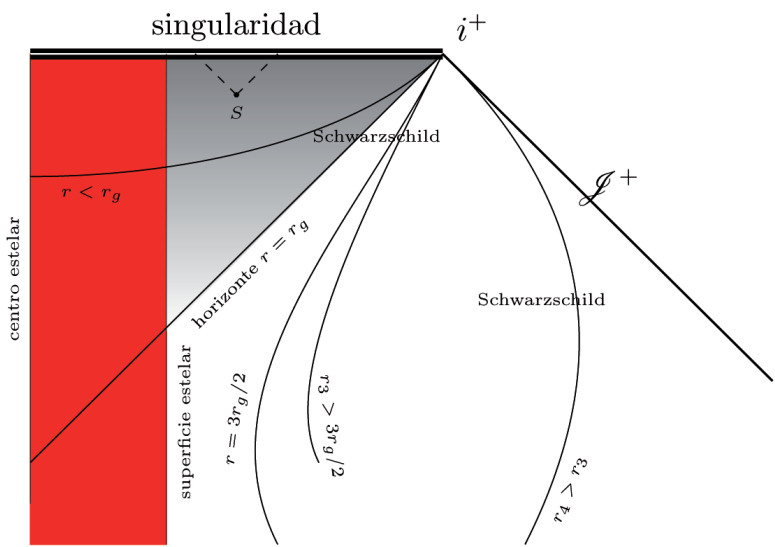

**Figura 4.** Un ejemplo de un diagrama de Penrose, correspondiente al colapso de la Figura 2. Cada punto representa una esfera (excepto en el centro estelar). La idea básica es compactificar el espacio-tiempo, trasladando el infinito a valores finitos de las variables *pero manteniendo los rayos luminosos radiales a 45° con la vertical.* La principal ventaja de esta idea es que se pueden visualizar las trayectorias *completas* de las partículas. Por ejemplo, aquellas que están en órbita con $r > r_g$ constante describen las líneas continuas mostradas y todas tienden a $i^+$, llamado *infinito temporal futuro.* Los rayos luminosos radiales llegan a $\mathscr{I}^+$, llamado *infinito luminoso futuro.* Hay, en general, otras tres zonas del infinito (no futuro) no mostradas. Las hipersuperficies con $r < r_g$ constante son espaciales, o sea, $r$ es una variable *temporal en el interior del HN.* En este diagrama se ve claramente que el horizonte es el borde del pasado casual de $\mathscr{I}^+$, y que los observadores externos *están sometidos al campo gravitatorio de la estrella antes de convertirse en HN.* Existe una órbita circular (inestable) para los fotones en $r = 3r_g/2$, y éste es el límite de la zona visible que puede aspirar a ver un observador lejano y el que produce la famosa silueta en la "foto" de un HN.

(3) La hipótesis del censor cósmico *débil* [22] es la afirmación de que todas las singularidades (o sea, incompletitudes) que se produzcan en el Universo quedan revestidas de su correspondiente horizonte aislándolas del mundo exterior. Aquí hay que exceptuar la singularidad inicial, que es obviamente visible, pero de otra índole. No hay resultados concluyentes al respecto hoy en día, y ciertamente hay tanto fervorosos adeptos como esforzados detractores, pero en todo caso lo cierto es que (i) la idea ha permitido desarrollos monumentales en RG y (ii) no se ha visto ninguna cosa que pueda considerarse una singularidad desnuda (Figura 5), mientras que hay miríadas de claros candidatos, si no ganadores, de hoyos negros. Muy en particular nuestro propio (probable) hoyo negro, Sgr A*, responsable de la otra mitad del Nobel de este año.

Hay una versión *fuerte* de la hipótesis del censor que asegura que ni siquiera infiltrándonos en el interior de un HN seremos capaces de observar la singularidad. Esto es precisamente lo que ocurre en el colapso de Oppenheimer-Snyder que la tiene *en el futuro* de cualquier observador, partícula o fotón que se pasee por allí (Figuras 2 y 4). Esta hipótesis se ha convertido en un problema matemático de primera magnitud y hay muchos resultados recientes al respecto aunque, se ha de conceder, son confusos [10].

En definitiva, Penrose fue el primero en tomarse en serio la posibilidad de que existieran los hoyos negros, los divulgó con entusiasmo y, sobre todo, con prodigiosa sabiduría. En [22] escribió:

Sólo deseo hacer un llamamiento para que se tomen en serio los "hoyos negros" y se exploren sus consecuencias con todo detalle. Porque, ¿quién puede decir, sin un estudio cuidadoso, que no puedan desempeñar un papel importante en la configuración de los fenómenos observados?

Hoy en día, ciertamente nadie puede atreverse ya a decirlo.

## Epílogo

Ignorando las sutilezas matemáticas y los pormenores de los teoremas, a veces se declara que las singularidades son *consustanciales* a la relatividad general. Nada más lejos de la realidad. Baste decir que todos los sistemas gravitatorios conocidos (sistema planetarios, estrellas, galaxias, etc.) son regulares y se describen eficazmente con la relatividad y sus límites de gravedad (post)newtoniana. Las únicas posibles excepciones son el Universo y el interior de los HN. En consecuencia una perspectiva más ajustada a la realidad es proclamar que *algunas singularidades* son un rasgo distintivo de la RG. Ofrecen indicios sólidos de la *necesidad de correcciones (posiblemente cuánticas) de la teoría en el interior de los hoyos negros.* En suma, los teoremas implican que la teoría es incompleta y debe ser corregida en situaciones extremas. ¡Maravilla es que la teoría muestre sus propias limitaciones!

## Comentarios finales

Permítanme para acabar que haga de portavoz alegre de la comunidad relativista. Como escribí en estas mismas páginas con ocasión del número monográfico [27] por el centenario de la teoría (que por cierto puede consultarse para ampliar muchas de las cosas aquí plasmadas), la RG fue el "patito feo" de la física durante décadas. Los lectores y colegas más j**óvenes pueden sorpren**derse con esta afirmación, pero el hecho es que la física gravitatoria relativista fue despreciada por los propios colegas de otras ramas durante mucho, demasiado, tiempo. A menudo con abierta animosidad. Para muestra, Bryce de Witt escribe en [28]:

La mayoría de ustedes no pueden hacerse idea de cuán hostil era la comunidad física [...] hacia las personas que estudiaban la RG [...]. A mediados de los [1950] Sam Goudsmit, entonces

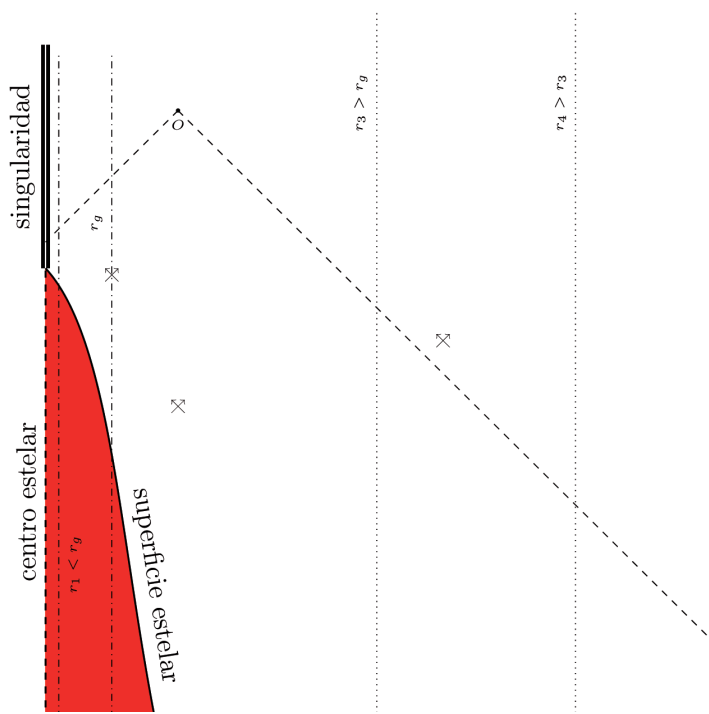

**Figura 5.** Como en la Figura 2, una representación más completa se obtiene rotando imaginariamente el dibujo en torno al eje definido por el centro estelar. Si el censor cósmico no actúa y no se forma un horizonte en el proceso de colapso, la situación sería como se muestra, con toda la estrella concentrándose en el punto central, produciendo una singularidad que, en este caso, es visible. Por ejemplo, una observadora en O ve la singularidad. Lo que es peor, la singularidad la puede afectar a ella. Por eso dar datos iniciales antes de que se forme la singularidad no determina la evolución del espacio-tiempo: la singularidad puede tener efectos que no están codificados inicialmente. La hipótesis del censor cósmico (no probada) prohíbe esta situación.

Labels within figure: singularidad; centro estelar; superficie estelar; $r_1 < r_g$; $r_1 < r_g$; $r_g$; $r_2 > r_g$; $r_3 > r_g$; $r_4 > r_g$

editor jefe de *Physical Review,* hizo saber que pronto aparecería un editorial diciendo que *Physical Review* y *Physical Review Letters* ya no aceptarían "artículos sobre gravitación [...]". Que este editorial no apareciera se debió a los esfuerzos entre bastidores de John Wheeler.

¡Afortunadamente! En [5] se atribuye al ingenio de Fred Hoyle la siguiente humorada, hablando del éxito que tuvo el congreso de Dallas:

aquí encontramos un caso que le permitía a uno sugerir que los relativistas con su trabajo sofisticado no sólo eran magníficos adornos culturales, sino que en realidad ¡podrían ser útiles para la ciencia!

La malquerencia hacia la RG y sus progresos no cesó hasta bastante después (si es que no sobreviven algunos rescoldos). Pero el tiempo, y la maravillosa física desarrollada por los relativistas, han puesto las cosas en su sitio. El premio Nobel 2020 ha sido el tercero en cuatro años para teóricos de RG y cosmología, con K. Thorne en 2017 y J. Peebles en 2019. El Nobel de este año, no obstante, tiene un sabor aún más dulce si cabe al haber premiado trabajo principalmente matemático. Un reconocimiento inequívoco a la relatividad matemática y sus *implicaciones físicas*.

Ha sido una muy grata sorpresa.

## Bibliografía

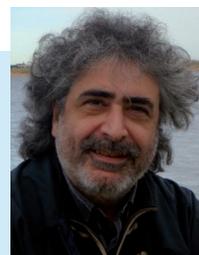

**José M. M. Senovilla**
Dpto. de Física
Universidad del País Vasco
UPV/EHU